\documentclass[useAMS,usenatbib]{mn2e}

\usepackage {graphicx}
\usepackage {times}

\title[Post-starburst--AGN Connection]{Post-starburst--AGN Connection: Spatially Resolved Spectroscopy of H$\delta$-Strong AGNs}
\author[T. Goto]
{Tomotsugu Goto$^{1}$\thanks{E-mail:tomo@ir.isas.jaxa.jp}
  \\
  $^{1}$ Institute of Space and Astronautical Science,  Japan Aerospace Exploration Agency,
 3-1-1 Yoshinodai, Sagamihara, Kanagawa 229-8510, Japan
}
\begin{document}
\date{\today; in original form 2005 November 21}
\maketitle


%






\begin{abstract}\label{abstract}
Ever since the co-existence of an active galactic nucleus (AGN) and a
 starburst was observationally discovered, there has been a significant
 controversy over whether there is a physical connection between
 starbursts and AGNs. If yes, it is a subject of interest to reveal
 which one triggers another.  Here we bring a unique insight onto the
 subject by identifying 840 galaxies with both a post-starburst
 signature (strong Balmer absorption lines) and an AGN (based on the
 emission line ratio).  These poststarburst-AGNs account for the 4.2\%
 of all the galaxies in a volume-limited sample.  The presence of a
 post-starburst phase with an active AGN itself is of importance,
 suggesting that AGNs may outlive starbursts in the starburst-AGN
 connection.  In addition, we have performed spatially resolved
 spectroscopy of three of our poststarbusrst-AGN galaxies, obtaining
 some evidence that the post-starburst region is more extended, but
 sharply centred around the central AGN, confirming a spatial
 connection between the post-starburst and the AGN.
\end{abstract}

\section{Introduction}\label{intro}

    It has been found that the nuclei of many nearby galaxies show signs of active galactic nuclei (AGN; in this paper AGN includes low-ionization nuclear emission regions or LINERs). 
    Evidence of starburst features have been detected in numerous AGNs
     \citep{1996ApJ...465...96N,1997ApJS..108..229H,1995ApJ...454...95M,1997ApJ...485..552M,2000A&A...357..850B,2004A&A...428..373B,1998MNRAS.297..579C,2001ApJ...558...81C,2004ApJ...605..105C,2005MNRAS.356..270C}. A QSO PG1700+518 shows a nearby starburst ring \citep{1999ApJ...512..140H}. Near infrared and CO mapping reveals a massive circumnucler starburst ring in I Zw 1 \citep{1998ApJ...500..147S}. A binary QSO FIRST J164311.3+315618B shows a starburst host galaxy spectrum \citep{1999ApJ...520L..87B}. HST-UV spectra revealed unambiguous signatures of massive stars (CIV and SiIV P Cygni lines) in Seyfert galaxies \citep{1997ApJ...482..114H,1998ApJ...505..174G}. Also, a starburst in Wolf-Rayet phase is detected in nearby Seyfert \citep{1997ApJ...482..114H,1998ApJ...501...94S,1999IAUS..193..703H,2001ApJ...546..845G}. 
      Continuum and emission line properties of AGNs are also found in
    many nearby galaxies both by optical surveys
    \citep{1992ApJ...393...90H,1997ApJS..112..315H,1999MNRAS.303..173S,2005AJ....129.1783H,2005AJ....129.1795H}
    and X-ray surveys \citep{1998A&A...330...25G}. Half of the
    ultraluminous infrared galaxies contain simultaneously an AGN and
    starburst activity \citep{1998ApJ...498..579G}. There also has been
    active debate whether ultraluminous infrared galaxies may evolve
    into QSOs \citep{1988ApJ...325...74S,1996ARA&A..34..749S,
    GotoIRAS}. Readers are referred to \citet{2003RMxAC..16..189G} and \citet{2004NuPhS.132..229R} for a review on the starburst/AGN indicators at various wavelengths.
      The discovery of tight correlation between black hole mass and bulge velocity dispersion provides another evidence that the formation of bulge and the central black hole may be closely linked \citep{2000ApJ...539L...9F,2000ApJ...539L..13G,2001MNRAS.324..757G,2004ApJ...613..109H}.   
 Thus, violent events of starburst and the AGN can coexist.
 To understand galaxy formation theory and the AGN mechanisms, it is important to characterize possible relationship between the starburst and AGNs. If there is a physical connection between starburst and AGNs, understanding the connection assists in creating unified models of AGNs. 
 

 In the theoretical perspective, there are currently two mainstreams to explain the connection between starbursts and AGNs. 
\begin{enumerate}
\item{\citet{1983ApJ...266..479W} has argued that starburst events in the nuclei of
     galaxies would evolve rapidly into compact configurations as
     dynamically distinct entities. The compact configuration of the
     nuclear-starburst collapses to form a single massive black hole
     generally favored as the central engine for AGNs \citep{1985Sci...230.1372M}. \citet{1993A&A...280..468V} suggested that starburst-induced turbulence in the interstellar medium (ISM) is responsible for the final stages of infall of gas into the AGN \citep[see also][]{2000ApJ...529..845C}. }
\item{On the other hand, \citet{1999A&AS..135..437G} suggest that it is the central AGN which gravitationally disturbs ISM, and thus, triggers circumnuclear star formation \citep[see also][]{1993ApJ...414..563V,1985ApJ...293...83V,1997ApJ...490..698D}.}
\end{enumerate}

 In summary, there has been a controversy over whether (1) starburst occurs first later to collapse to AGNs; or (2) AGN forms first fueling starburst activity later.  Both of these theories assume that there is a connection between the starburst and AGN. 

 However observationally, evidence for either case is fragmented at present. 
 It has been simply difficult to trace the complicated process of the galaxy evolution and the AGN formation with Giga
 years of timescale, using the observation which only brings us with
   information of a single epoch. 
 In this connection, it is very important to study starbursting AGNs, in which the transition between AGNs and starburst activities are currently on-going.

 With the advent of the Sloan Digital Sky Survey (SDSS), we have a chance to obtain a unique perspective on the subject. Out of  $\sim$210,000 spectroscopic galaxies in the SDSS Data Release 2 \citep[SDSS DR2; ][]{2004AJ....128..502A} with enough signal-to-noise ratio,  we have identified 840 galaxies with both strong H$\delta$ absorption \citep[and thus, in the poststarburst phase; see][]{2003PhDT.........2G,2004A&A...427..125G} and the emission line ratio consistent to be an AGN. All of our sample galaxies are in the strong post-starburst phase with H$\delta$ equivalen width of greater than 4\AA\, which have been difficult to be found previously. These galaxies can bring us a new view on the subject, in which the starburst activity is ceasing whereas the AGN is still active. In this paper, we further perform spatially resolved spectroscopy of three of these galaxies, in order to shed light on the spatial connection between the post-starburst and the AGN. 
   Unless otherwise stated, we adopt the best-fit WMAP cosmology: $(h,\Omega_m,\Omega_L) = (0.71,0.27,0.73)$ \citep{2003ApJS..148....1B}.

\section{Sample Selection}\label{data}

 We have selected our sample from the second data release of the SDSS \citep{2004AJ....128..502A}. 
  We only use those objects classified as galaxies ({\tt type=3}, i.e., extended) with spectroscopically measured redshift of $z>0.01$. These two criteria can almost entirely eliminate contamination from mis-classified stars and HII regions in nearby galaxies. We further restrict our sample to those with spectroscopic signal-to-noise $>$10 per pixel in the continuum of the $r$-band wavelength range since it is difficult to measure weak absorption lines when the signal-to-noise is smaller than 10 \citep[see][]{Goto2003ea}. 
 We, then, have selected post-starburst galaxies with AGN activity as galaxies with strong Balmer absorption lines, and emission line ratios consistent to be AGNs.  Strong Balmer absorption lines such as H$\delta$, H$\gamma$ and H$\beta$ show that the galaxy spectra is dominated by A-type stars with the absence of massive O,B-type stars. Since the lifetime of A-stars is about one Giga year, these galaxies have experienced starburst within the last one Giga year and stopped it by now, i.e., they are in the post-starburst phase \citep{1983ApJ...270....7D,1999ApJ...518..576P,2004A&A...427..125G}. 
We select post-starbursts as galaxies with  H$\delta$ equivalent width (EW) greater than 4\AA\ in absorption.  The H$\delta$ line is preferred over other hydrogen Balmer lines (e.g., H$\epsilon$, H$\zeta$, H$\gamma$, H$\beta$)
 since the line is isolated from other emission and absorption lines, as well
 as strong continuum features in the galaxy spectrum (e.g.,
 D4000). Furthermore, the lower order Balmer lines (H$\gamma$ and H$\beta$)
 can suffer from significant emission-filling, while the higher order lines (H$\epsilon$ and H$\zeta$) have a low
 signal-to-noise in spectra. 
 We have adopted the flux-summing technique used in \citet{Goto2005266ea} in measuring H$\delta$ EW. 
 We select those galaxies with H$\delta$ EW $>4$\AA\ in absorption as poststarbursts.
We stress that our criteria of H$\delta$ EW $>4$\AA\ is a very strong one compared with previous work.
 Thus, we can select strong post-starburst galaxies, which previous work with a smaller mother sample were not able to find \citep[e.g.,][]{2003MNRAS.339..772R,2005MNRAS.356..270C}. 

 We regard a galaxy as an AGN when a galaxy satisfies the [OIII]/H$\beta$ and [SII]/H$\alpha$ flux ratio proposed by \citet{2001ApJ...556..121K}. Note that this is more conservative criteria in selecting AGNs than was used in \citet{2003MNRAS.346.1055K}, i.e., only those galaxies that cannot be explained by any starburst model are selected as AGNs. These criteria are similar to the one used in \citet{2003ApJ...597..142M}. Out of 209,938 galaxy spectra which satisfies the S/N and redshift criteria  in the SDSS DR2, 14294 galaxies have AGN signatures. Among them, 840 galaxies have both the post-starburst and AGN signatures (We dub them as H$\delta$-strong AGNs or HDSAGNs hereafter).  Previously, only a few of post-starburst+AGN galaxies were known \citep[e.g.,][]{1999ApJ...520L..87B,2000MNRAS.318..309D}. We emphasize the strong post-starburst features in our sample galaxies (H$\delta$ EW$>$4\AA). Previously, it has been known that some AGNs have higher order Balmer absorption lines \citep[e.g.,][]{2001ApJ...546..845G}. However, it has been difficult to select such strong absorption (H$\delta$ EW$>$4\AA) coexisting with another rare occasions of AGNs, simply due to the smaller mother sample size.  The large data set of the SDSS has enabled us to study a unique phase in the starburst-AGN connection where the galaxy is so close to the epoch of the last star formation ($<$1 Gyr). 
Since our galaxies are close to the epoch of the truncation of the starburst, 
 it is  expected that our sample potentially provide with more hints on the physical reason why these galaxies shut-off starbursts, and on the physical connection with their AGN activity.

\begin{table*}
\caption{Three HDSAGNs observed with the APO 3.5m telescope.
}\label{tab:targets}
\begin{center}
\begin{tabular}{crrrrrrrrl}
\hline
 Name   &   RA & DEC & redshift & $r$ & $R_{90}$($r$; arcsec) & kpc/arcsec & H$\delta$ EW (\AA) & Observing Date & Exposure time (min) \\
\hline
\hline
SDSSJ023301.2+002515 &  02:33:01.24 & +00:25:15.0 & 0.0225 &  13.79 &  16.06 & 0.45 & 5.40$\pm$0.17 & Dec. 17, 2004 & 30 \\
SDSSJ081347.5+494110 &  08:13:47.49 & +49:41:09.7 & 0.0939 &  17.23 &   3.68 & 1.72 & 6.12$\pm$0.58 & Nov. 11, 2004 & 60 \\
SDSSJ014154.9+141133 &  01:41:54.88 & +14:11:33.0 & 0.0958 &  17.61 &   9.36 & 1.77 & 4.18$\pm$0.25 & Nov. 11, 2004 & 60 \\
\hline
\end{tabular}
\end{center}
 \end{table*}

\section{Spatially resolved spectroscopy of three HDSAGNs}

Although the SDSS spectra lead us to an interesting discovery of the post-starburst AGNs, the SDSS spectra do not contain spatial information due to the limited aperture size of 3 arcsec of the SDSS fiber spectrograph.
In order to understand spatial connection between the AGN and post-starburst,   we have performed a spatially resolved long-slit spectroscopy of three of them using the Apache Point Observatory (APO) 3.5m telescope. This observation can  potentially reveal spatial connection between the post-starburst and the AGN.
Among the 840 H$\delta$-strong AGNs (HDSAGNs, hereafter) selected in Section \ref{data}, three galaxies that were visible on the observing dates were observed with the Dual Imaging Spectrograph (DIS) mounted on the APO 3.5m telescope in the long-slit mode with 1.5'' slit-width. We used the blue low and red medium resolution, corresponding to the resolution of 2.43\AA/pix and 2.26\AA/pix for the blue and red camera, respectively.  Spacial scales for the blue and red gratings are 0.42 and 0.40 arcsec/pixel, respectively.

 Observations were carried out on the nights of November 11th and December 17th, 2004. We present the basic information of the three targets in table \ref{tab:targets}, in which measured quantities such as positions, redshift, and Petrosian radius are taken from the SDSS catalog. Here, $R_{90}$ is the radius within which 90\% of the $r$-band Petrosian flux is contained. Magnitudes are de-reddened Petrosian magnitude in $r$-band. The physical scale based on the WMAP cosmology is shown in the unit of kpc/arcsec for convenience. 
The seeing was $\sim$0.9 arcsec on both nights based on the PSF measured with standard stars. Data reduction was carried out using the standard IRAF routines.

 In the panel (a) of Figures \ref{fig:SDSSJ023301.2+002515}-\ref{fig:SDSSJ014154.9+141133}, we show the $g,r,i$-composite images of the three targets. In the panel (b) of Figures \ref{fig:SDSSJ023301.2+002515}-\ref{fig:SDSSJ014154.9+141133}, we present spectra of the targets taken with the SDSS 3'' fiber spectrograph. The spectra are shifted to the rest-frame and smoothed using a 20 \AA\ box. Note that all three targets have strong emission lines whose flux ratios are consistent to be an AGN. At the same time, they have strong Balmer absorption lines (such as H$\delta$) characteristic to the post-starburst phase. We repeat that it has been almost impossible to find HDSAGNs in such strong poststarburst phase without the large mother sample of the SDSS, nicely differentiating this work on the poststarburst-AGNs from previous work on the starbursting-AGNs \citep[e.g., ][]{2003MNRAS.339..772R,2005MNRAS.356..270C}. The mysterious co-existence of the post-starburst and AGN in these three galaxies is clear even just by looking at these spectra in Figures \ref{fig:SDSSJ023301.2+002515}-\ref{fig:SDSSJ014154.9+141133}.





\section{Results}\label{sec:Results}
 
 In order to investigate spatial variation of the spectra of HDSAGN galaxies, we have divided each spectrum into 11 different bins in spatial direction along the slit. Since we sampled $\sim$20 arcsec along the slit (depending on the galaxy), one bin samples $\sim$2 pixels, which is close to the seeing size ($\sim$0.9 arcsec). On each spectrum, we measure H$\delta$, H$\beta$, [OIII], H$\alpha$, and [SII] lines including their errors using the flux summing technique described in \citet{Goto2003ea,Goto2005266ea}. Errors of the flux are measured based on the 1-$\sigma$ fluctuation in the continuum region around each line (50\AA\ regions on both blue and red sides of the line). When we take a ratio of two lines, these errors are propagated accordingly.

\subsection{SDSSJ023301.2+002515}\label{sec:SDSSJ023301.2+002515}
 In the panel (c) of Figure \ref{fig:SDSSJ023301.2+002515}, we
 compare spatial flux distribution of [OIII] emission line (dashed lines) and H$\delta$ absorption line (solid lines) of SDSSJ023301.2+002515. Both of the flux are normalized at the peak position to ease the comparison. The deficit of the H$\delta$ is computed based on the absorbed amount in the same manner as a flux is computed for  emission lines. Both of the fluxes are strongly peaked at the centre, but interestingly, the H$\delta$ deficit is slightly more extended than the [OIII] flux, especially at radius greater than $\pm$2 arcsec, or $\pm$1 kpc. The trend is clearer when EWs are compared in the panel (d). The [OIII] emission has large EWs of 6-12\AA\ only within 1 kpc. On the other hand, the H$\delta$ EWs stay large as far as $\pm$4 arcsec, or $\pm$2 kpc. In this panel, [OIII] EWs are positive for emission and H$\delta$ EWs  are positive for absorption. These two middle panels (c,d) suggest that the post-starburst region in this galaxy may be more extended than the emission line region. 
 
 It is interesting to see H$\beta$ EW profile of this galaxy in the panel (e) of Figure \ref{fig:SDSSJ023301.2+002515}, where the H$\beta$ EW is positive when the line is in absorption. 
  The H$\beta$ line is in absorption at the distance of  $>$1 kpc, perhaps corresponding to the post-starburst region. However, within the central $\pm$1 kpc, the H$\beta$ EW decreases to zero, perhaps due to the cancellation by the H$\beta$ emission in the central regions where [OIII] emission was also strong. 

 In the panel (f), we show the [SII]/H$\alpha$ flux ratio as a function of distance from the galaxy centre. The [SII]/H$\alpha$ flux ratio is peaked at the galaxy centre, and then decreases toward outside of the galaxy. It becomes especially lower at $>$2 arcsec, or $>$1 kpc. We note that the panels (f) in Figs. \ref{fig:SDSSJ023301.2+002515}-\ref{fig:SDSSJ014154.9+141133} often have fewer points since we do not measure [SII]/H$\alpha$ ratio when the H$\alpha$ line is in absorption and/or the [SII] line is not detected.

 In summary, the panels in Figure \ref{fig:SDSSJ023301.2+002515} suggest that both of the AGN and post-starburst regions in SDSSJ023301.2+002515 is centrally concentrated. However, the post-starburst region is more extended toward outside of the galaxy than the emission line region.



\subsection{SDSSJ081347.5+494110}\label{sec:SDSSJ081347.5+494110}

 In Figure \ref{fig:SDSSJ081347.5+494110}, we show the results for SDSSJ081347.5+494110. It is worth emphasizing that this galaxy has large H$\delta$ EW of 6.12$\pm$0.58 \AA\ within the 3 arcsec fiber of the SDSS (Table \ref{tab:targets}). Such a strong poststarburst-AGN is impossible to find without a large mother sample such as the SDSS ($\sim$210,000 galaxies were used in this work). Note that this galaxy has two peaks (cores) in the $g,r,i$-composite image in the panel (a). The Petrosian 90\% radius ($R_{90}$) was measured only for the northern peak, and is smaller than the actual spatial extent of the galaxy. 
 The long slit was oriented along these two peaks during the observation. 
 The comparison between [OIII] and H$\delta$ deficit in the panel (c) and EWs in the panel (d) shows a similar trend as was observed in Figure \ref{fig:SDSSJ023301.2+002515}, i.e., the [OIII] emission is strongly peaked at 0-2 arcsec, or 0-3.4 kpc, and immediately decreases outside of this region. On the other hand, the H$\delta$ absorption is more extended toward -4 arcsec, or -6.8 kpc. 
 H$\beta$ EW in the  panel (e) also shows strong absorption at around 0-2 arcsec, being consistent with the H$\delta$ absorption, although the  H$\beta$ absorption is not so spatially extended as the H$\delta$ absorption. 
  In the panel (f), the [SII]/H$\alpha$ flux ratio also shows a sharp peak at around 2 arcsec, corresponding the peaks of the other lines.  
 In the panel (c),  the H$\delta$ absorption has two peaks one at $\sim$+2 and one at 0 arcsec. These two peaks are both from the upper peak in the panel (a).  The H$\delta$ absorption is slightly weaker at $\sim$+1 arcsec, where [OIII] emission is peaked, perhaps due to the H$\delta$ emission filling from the central AGN.

 In summary, the previous picture where the post-starburst region shares a sharp spatial peak with the emission line region, but is more extended than the emission line regions holds true for this galaxy. This galaxy, however, has two peaks in the image in the panel (a) and the long-slit is aligned with these two peaks. Therefore, it is likely that one peak (upper in the figure) has both an AGN and post-starburst component, and only the post-starburst region extends to the other peak (lower left in the figure).

\subsection{SDSSJ014154.9+141133}\label{sec:SDSSJ014154.9+141133}
 
 Figure \ref{fig:SDSSJ014154.9+141133} shows the results for
 SDSSJ014154.9+141133 in a similar manner as previous two galaxies. The
 results of the comparison between [OIII] and H$\delta$ lines are very
 similar to that in Section \ref{sec:SDSSJ023301.2+002515}. In the panel
 (c), [OIII] flux is centred at $\pm$2 arcsec (or 3.5 kpc), whereas
 H$\delta$ deficit is more extended to +3-4 arcsec. The trend can also
 be seen in the panel (d) where H$\delta$ EWs shows high value ($>4$\AA)
 outside of $\pm$2 arcsec, whereas [OIII] is only strong at the
 centre. H$\beta$ EWs in the panel (e) does not show much trend, but is
 consistent with the H$\delta$ EW distribution in the panel (d). The
 [SII]/H$\alpha$ ratio does not vary either except a very low value at
 $\sim$+4 arcsec. At 4 arcsec away from the centre, however, the flux is
 relatively weak and a slight change in either [SII] or H$\alpha$ can change the ratio dramatically. Therefore, we do not strongly interpret the low value at  $\sim$+4 arcsec.  
 Overall, the trend observed with previous two galaxies can also be seen with SDSSJ014154.9+141133, i.e., both the AGN and the post-starburst is centrally peaked, but the post-starburst is more extended than the AGN region.
 We note that due to the fainter magnitude, the signal-to-noise ratio is lower for the latter two galaxies than for the SDSSJ023301.2+002515. 
 The signal-to-noise ratio of the blue spectra measured around  4750\AA\, where the spectrum is relatively flat, varies from 5 to 10 for SDSSJ081347.5+494110 and 5-23 for SDSSJ014154.9+141133 (highest at the centre of the galaxy). The signal-to-noise ratio for  SDSSJ023301.2+002515 varies from 20-41, depending on the spatial position of the galaxy.



\subsection{PSF deconvolution}\label{sec:deconvolution}

 In the previous subsections, we have obtained some evidence that the post-starburst regions may be more extended than the central emission line regions powered by AGNs. However, the spatial profiles in the previous section have been smeared by the seeing ($\sim$0.9 arcsec on both nights). In this subsection, we attempt to deconvolve the PSF from the profile in order to investigate the underlying true spatial profile of the HDSAGNs. 

 In de-convolving the PSF, we adopt the methodology used in \citet{YagiGoto}. We briefly describe the procedure here. Readers are referred to the reference for more details.
 First, we assume that spatial line profiles can be well-represented by the S\'ersic profile \citep{1963BAAA....6...41S,2005PASA...22..118G}. Since we normalize the flux intensity, free parameters here are $n$ (shape of the profile) and $R_e$ (effective radius; normalization in the radial direction). Then we convolve the S\'ersic profile with a PSF using 0.9 arcsec of Gaussian to this profile. After normalizing the flux using the peak intensity, we search for the least chi-square best-fit by changing the free parameters, $n$ and $R_e$. The S\'ersic profile with the best-fit parameters is the PSF-deconvolved galaxy profile we desire. 

 We perform this deconvolution for H$\delta$ and [OIII] flux profile for the three HDSAGNs. Results are shown in Table \ref{tab:abr}. For both lines of all the galaxies, $n$ is smaller than 4, reflecting the profiles of HDSAGNs are spiral-like rather than elliptical-like.
 Interestingly, the effective radius, $R_e$, is much larger for H$\delta$ than for [OIII] in all the three HDSAGNs. Especially, SDSSJ014154.9+141133 and SDSSJ081347.5+494110 have $R_e$(H$\delta$) more than three times as large as $R_e$([OIII]). The $R_e$(H$\delta$) is also larger than $R_e$([OIII]) for SDSSJ023301.2+002515. 
 These results support our finding in the previous subsections that the post-starburst region may be more extended than the AGN region, even after the seeing effects are removed.


\begin{table*}
\caption{
Best-fit parameters of deconvolved S\'ersic profile fit
}\label{tab:abr}
\begin{center}
\begin{tabular}{crrrr}
\hline
 Name   &   $n$(H$\delta$) & $R_e$(H$\delta$;kpc) & $n$([OIII]) & $R_e$([OIII];kpc) \\
\hline
\hline
SDSSJ023301.2+002515 &   1.8 &   1.6 &   0.8 &   1.1 \\ 
SDSSJ081347.5+494110 &   0.1 &  19.7 &   0.8 &   6.2 \\ 
SDSSJ014154.9+141133 &   2.6 &  20.2 &   1.0 &   3.5 \\ 
\hline
\end{tabular}
\end{center}
\end{table*}

\begin{figure*}
\begin{center}
\includegraphics[scale=0.35]{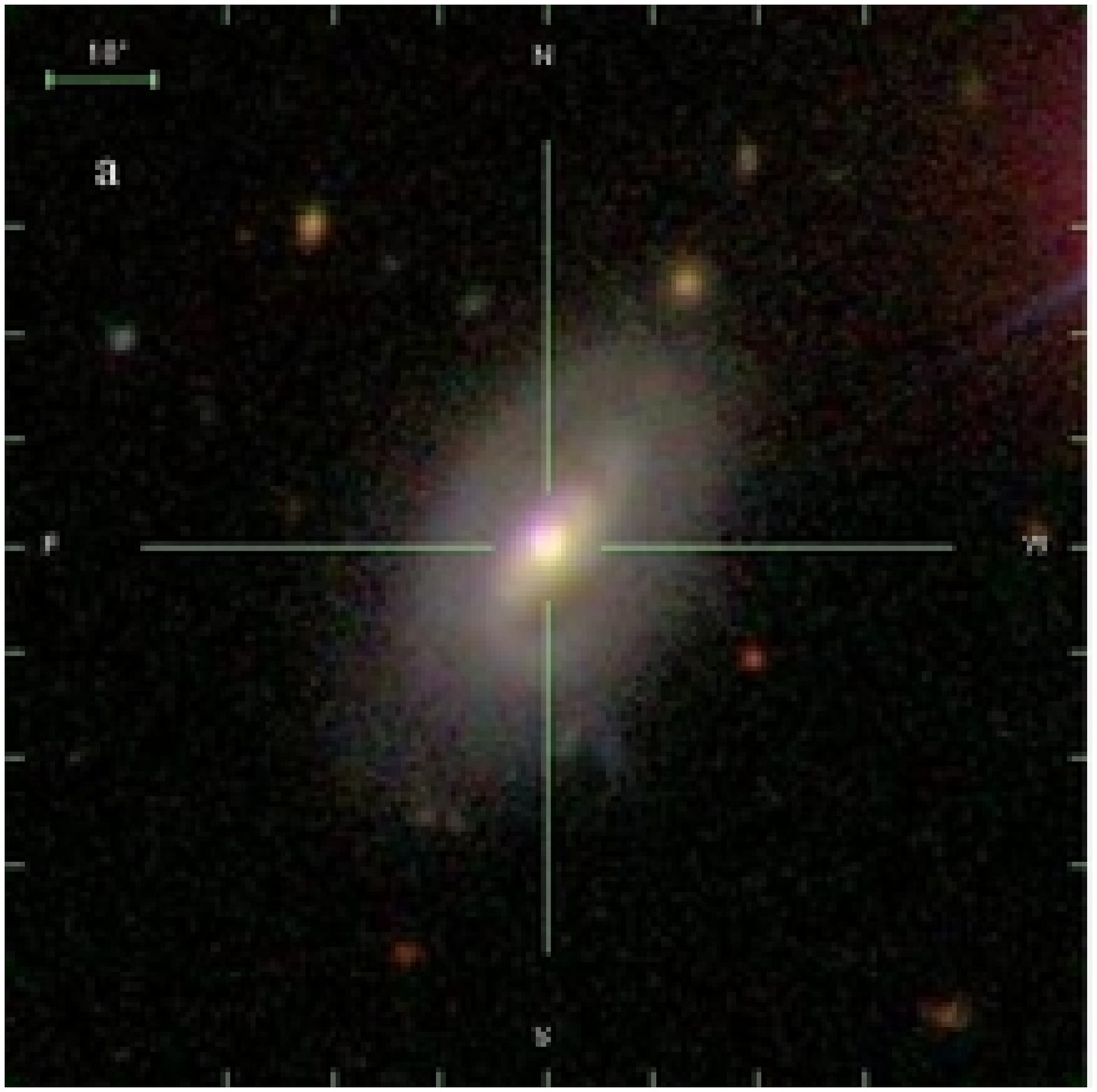}
\includegraphics[scale=0.45]{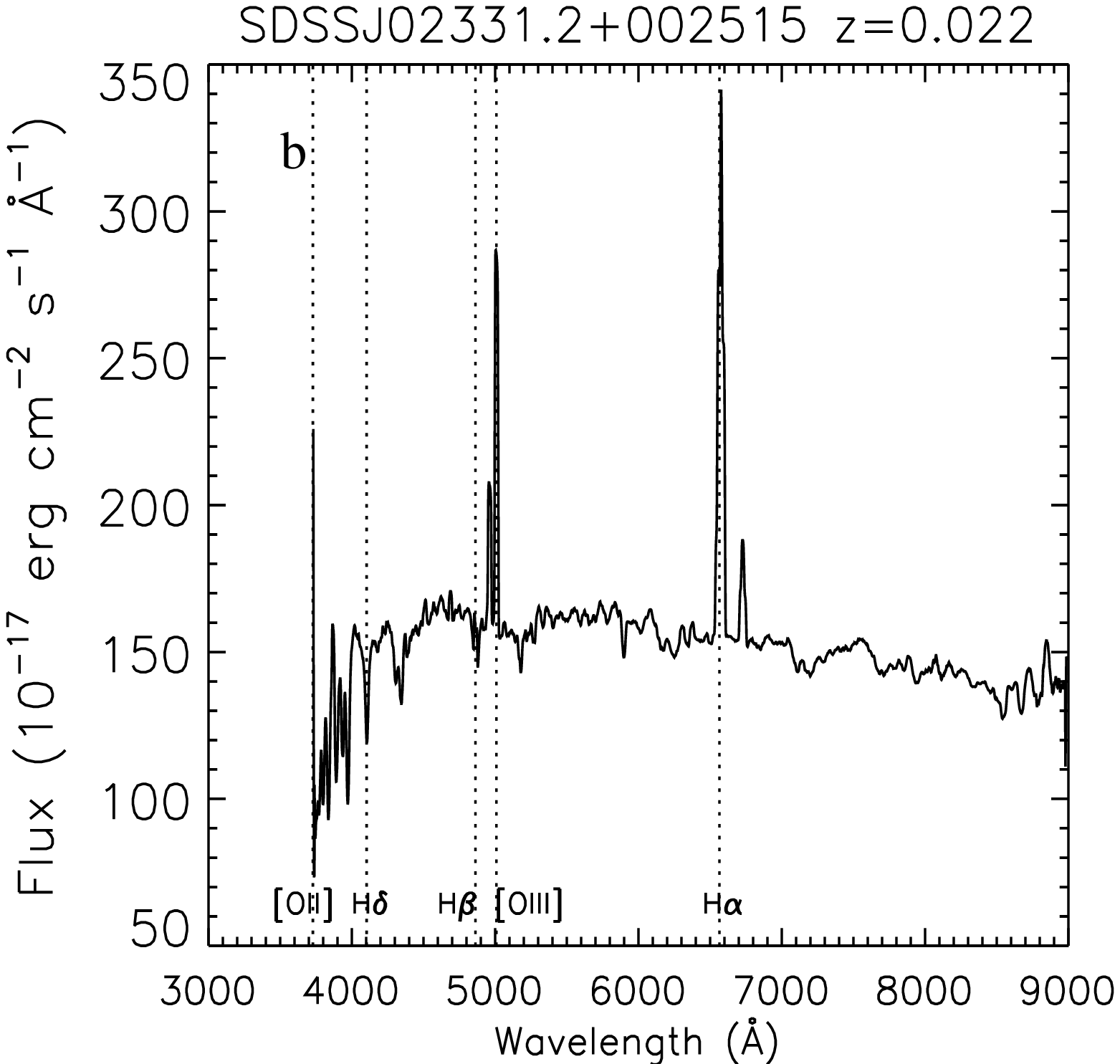}
\includegraphics[scale=0.40]{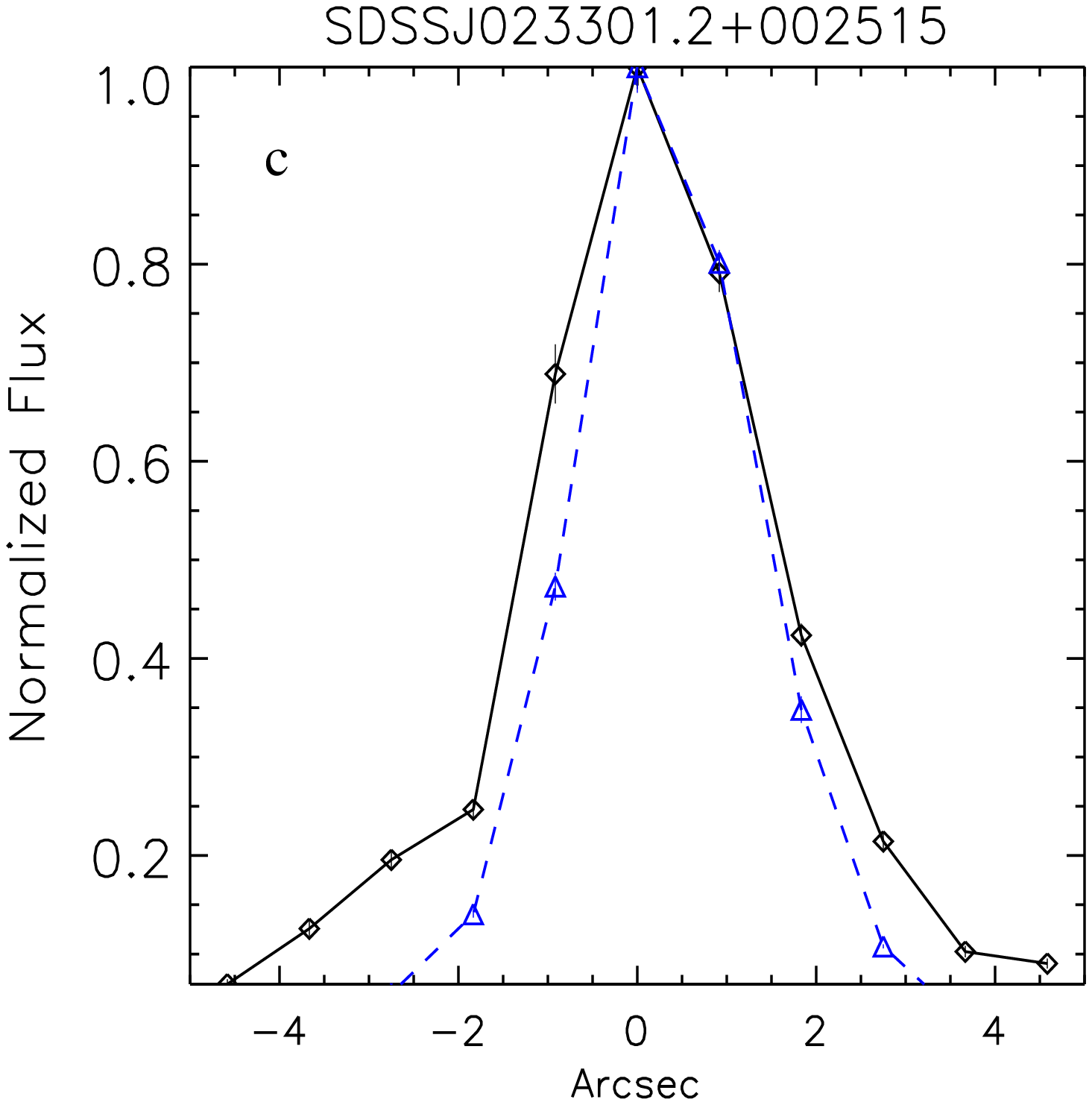}
\includegraphics[scale=0.40]{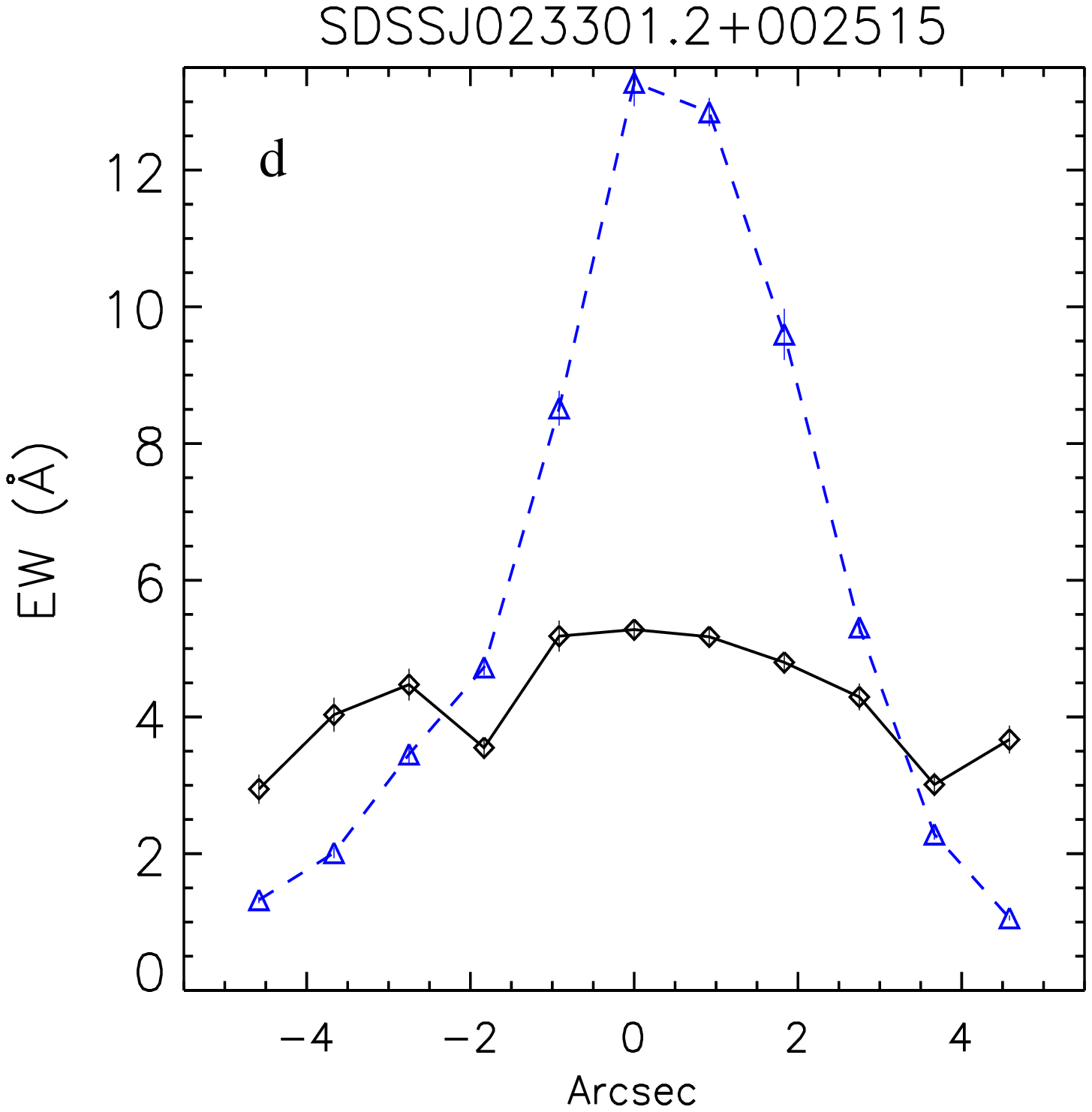}
\includegraphics[scale=0.40]{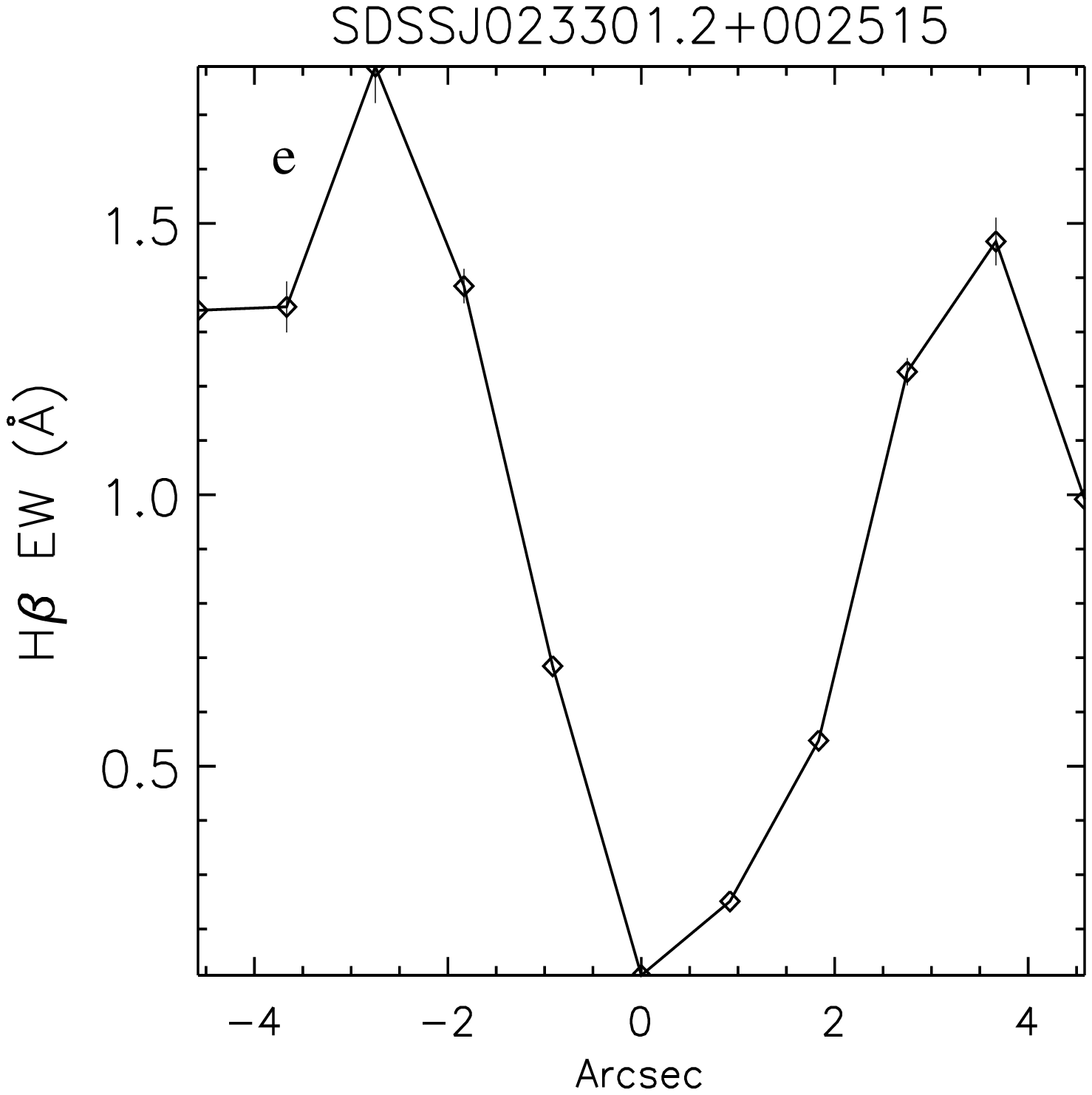}
\includegraphics[scale=0.40]{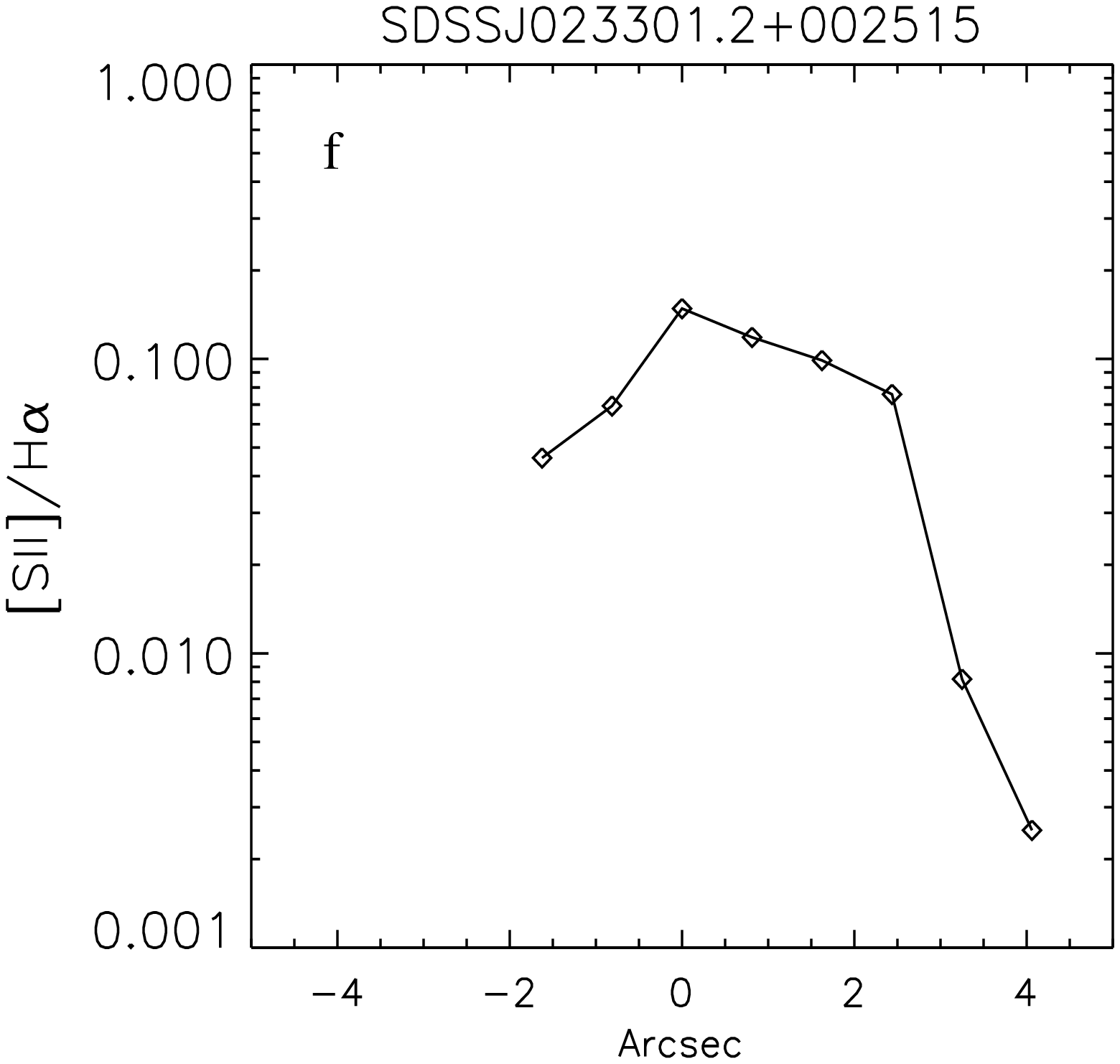}
\end{center}
\caption{Figures for SDSSJ023301.2+002515. 
 The panel (a) shows $g,r,i$-composite image taken with the SDSS. 
 The panel (b) shows the spectrum taken with the SDSS 3'' fiber spectrograph.
 Two middle panels (c,d) compare equivalent width (EWs) and flux normalized at the centre between [OIII] (dashed lines) and H$\delta$ (solid lines) as a function of spatial position on the slit.  In the panel (e), H$\beta$ EWs are shown as a function of spatial position.  Here, emission is positive for [OIII], whereas absorption is positive for H$\delta$ and H$\beta$.  The panel (f) shows the flux ratio of [SII]/H$\alpha$ as a function of spatial position.
}\label{fig:SDSSJ023301.2+002515}
\end{figure*}



\begin{figure*}
\begin{center}
\includegraphics[scale=0.45]{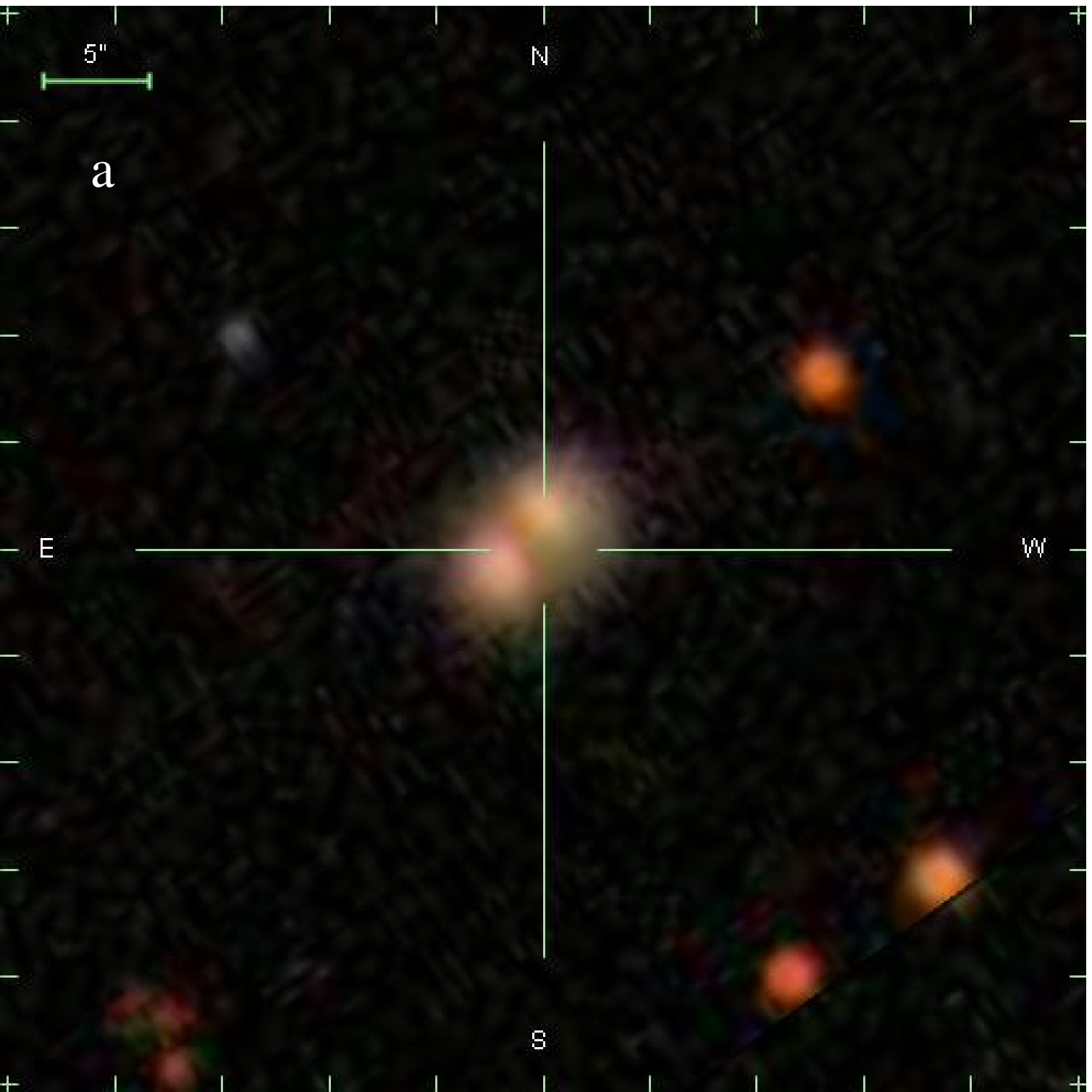}
\includegraphics[scale=0.45]{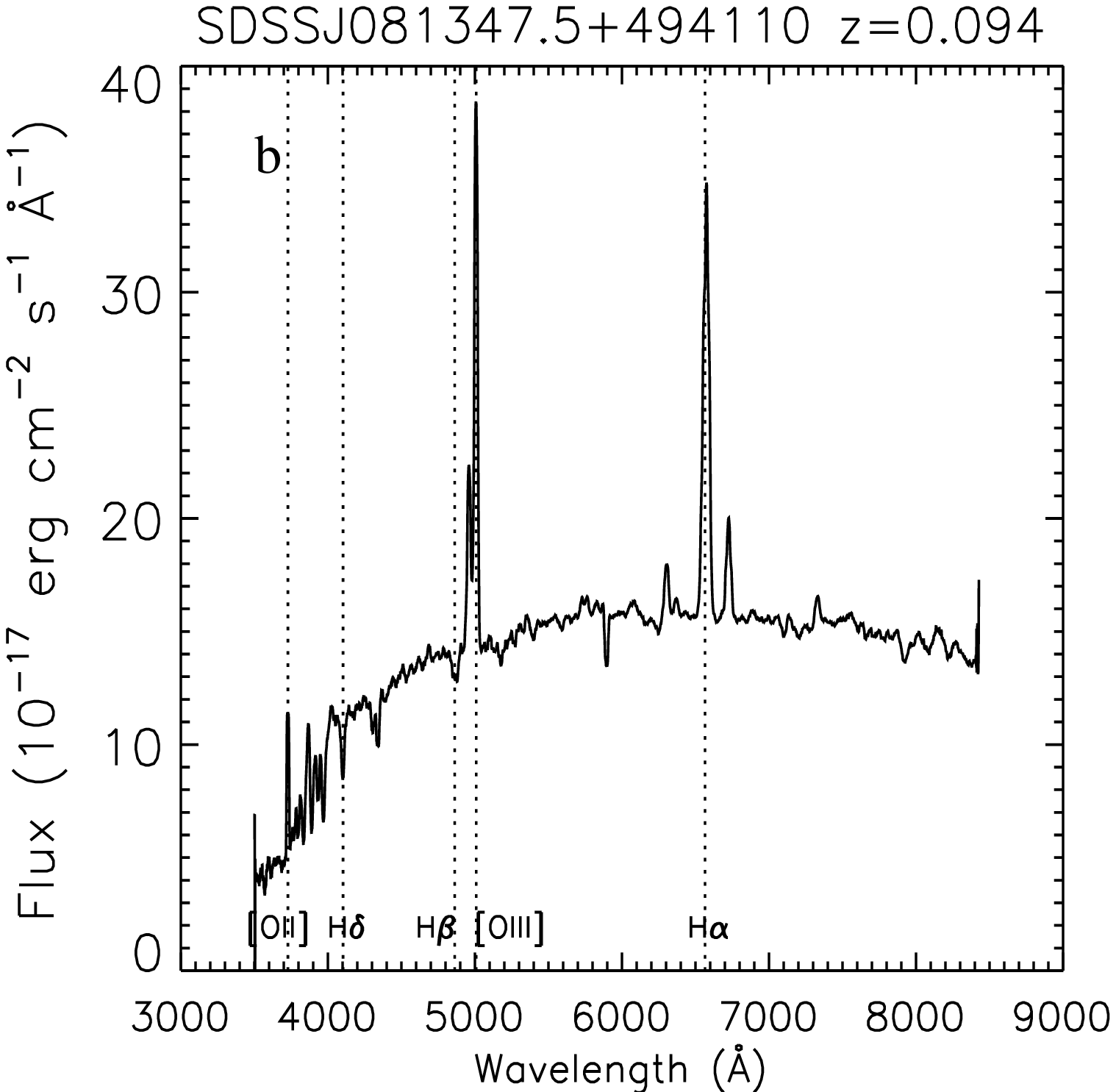}
\includegraphics[scale=0.40]{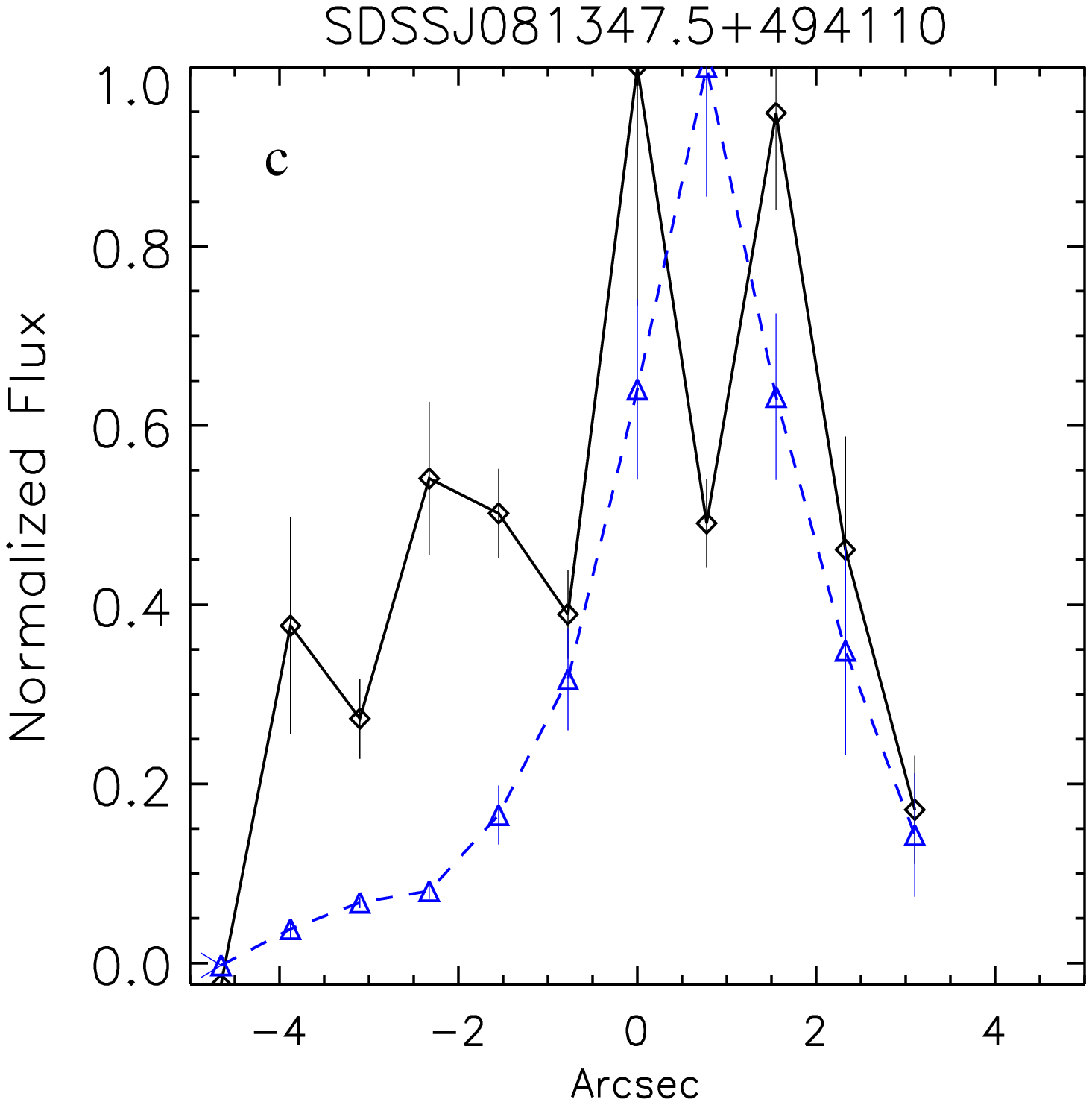}
\includegraphics[scale=0.40]{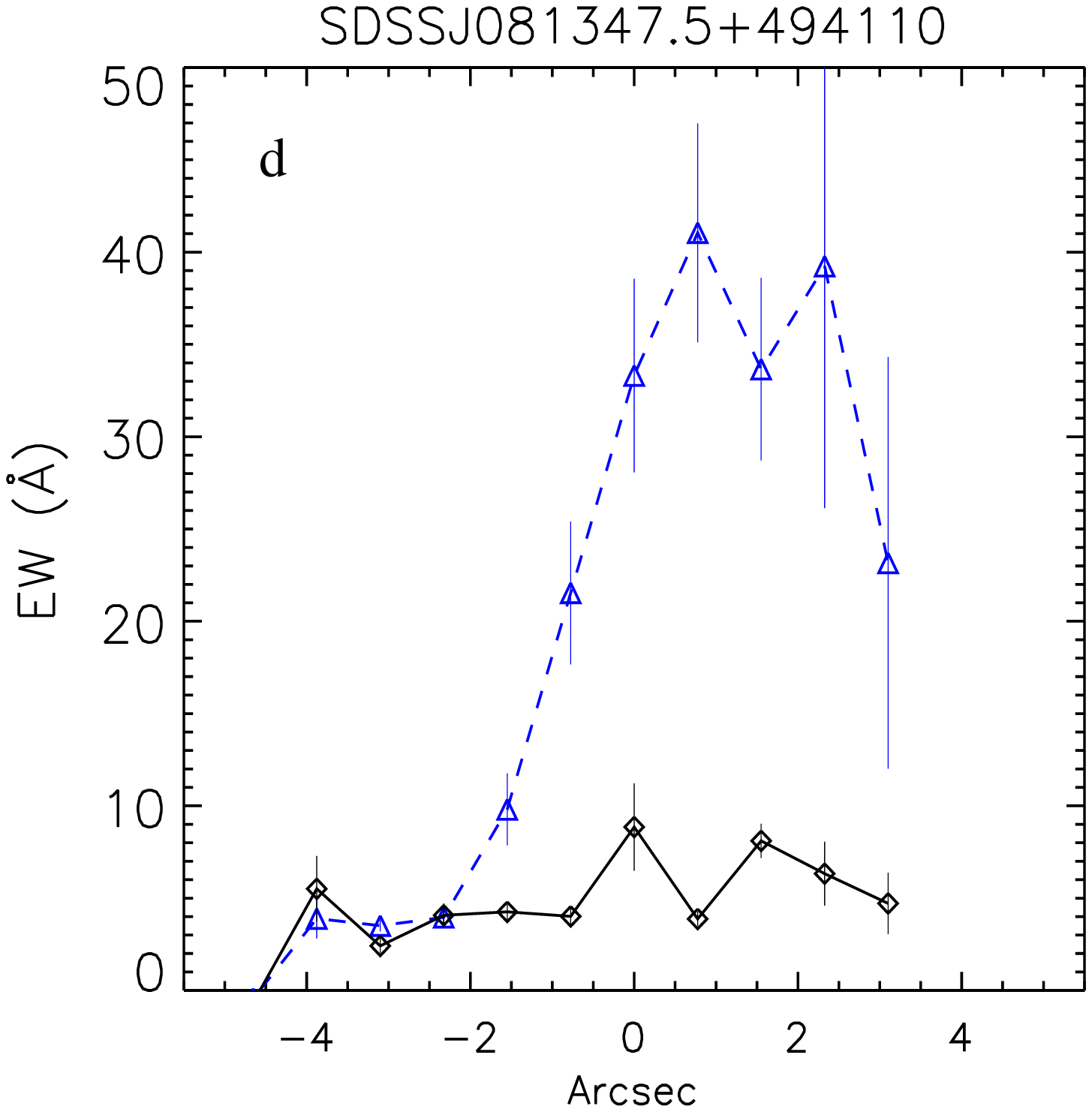}
\includegraphics[scale=0.40]{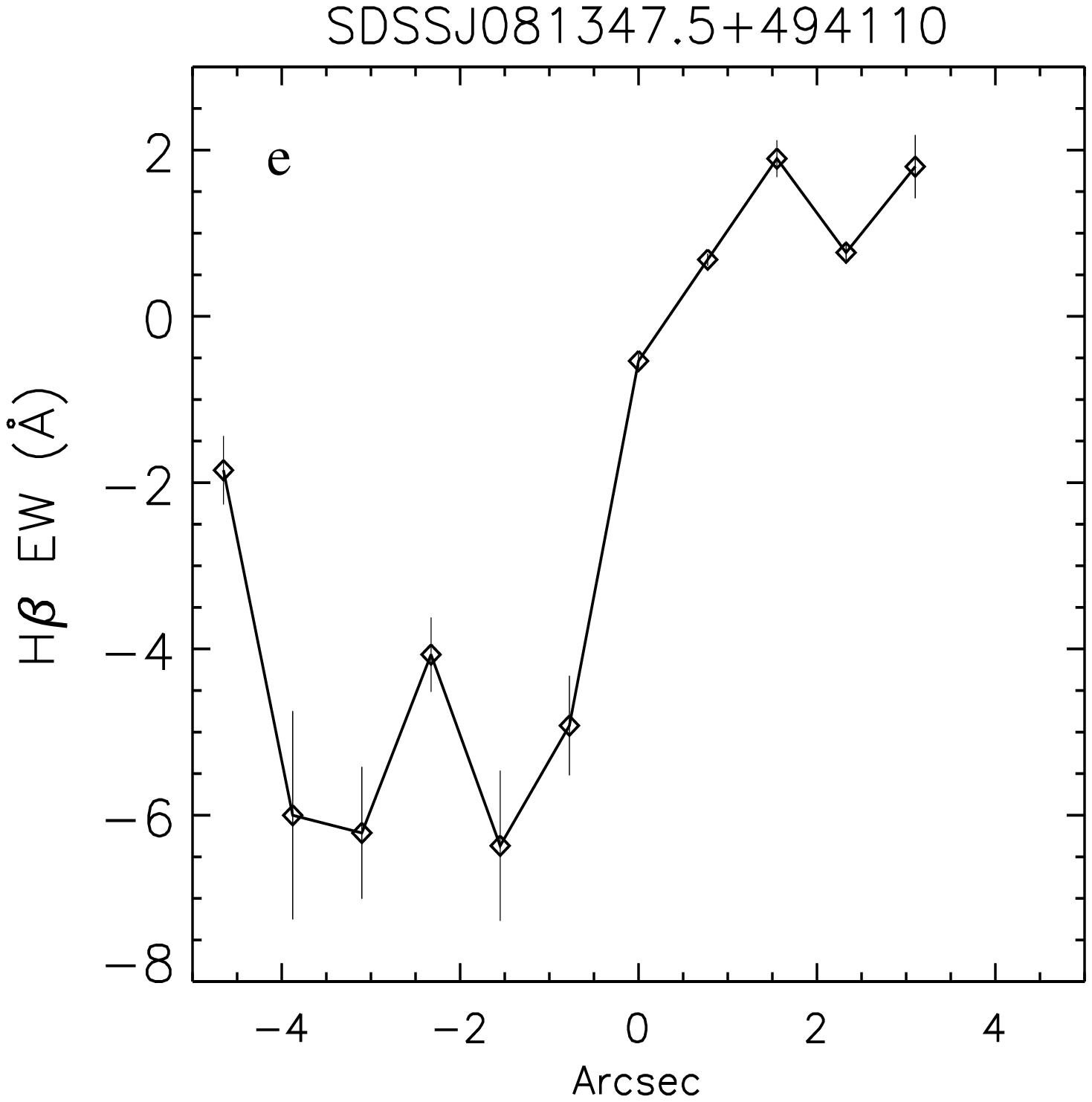}
\includegraphics[scale=0.40]{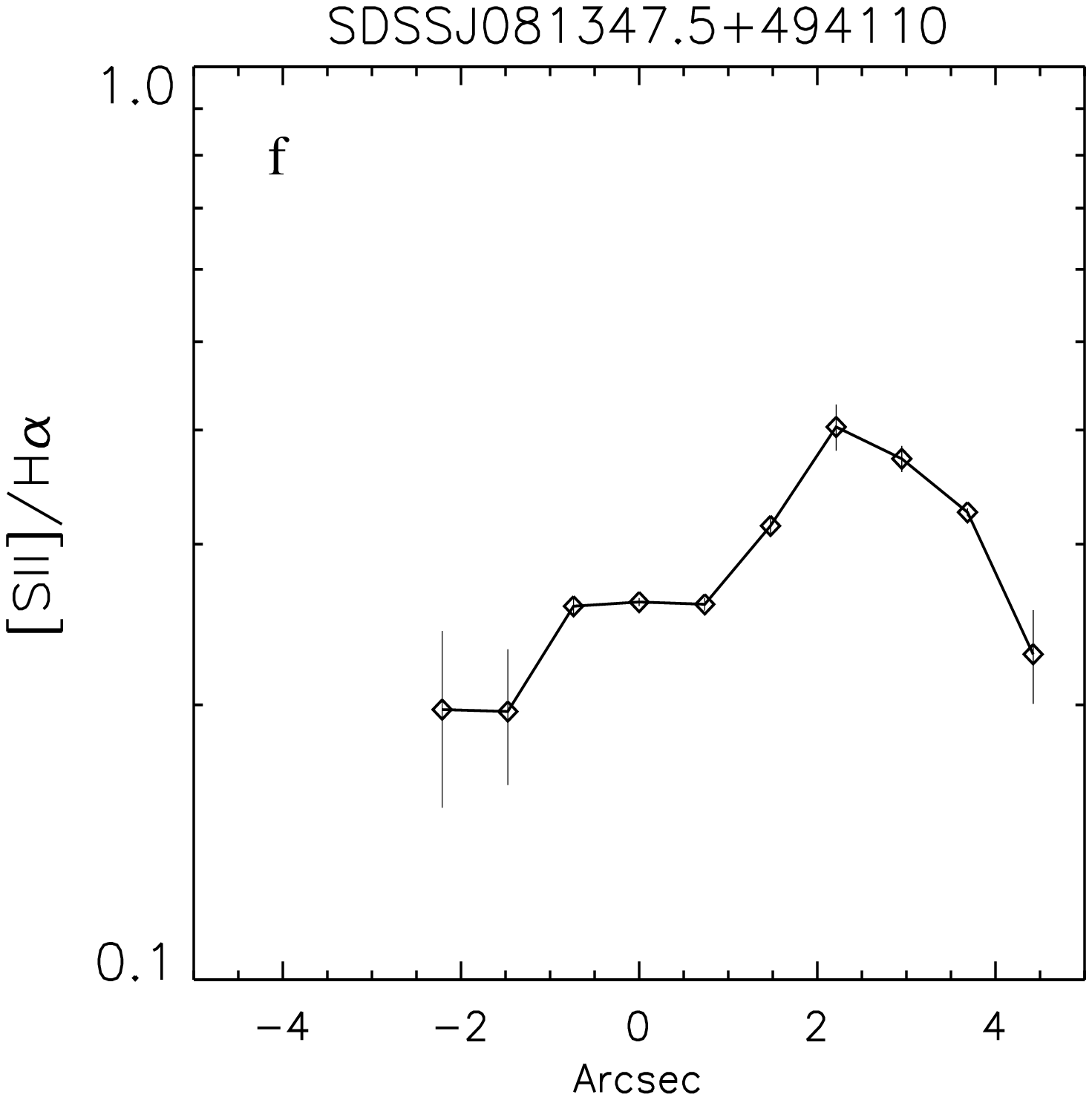}
\end{center}
\caption{As figure \ref{fig:SDSSJ023301.2+002515}, but for SDSSJ081347.5+494110. 
}\label{fig:SDSSJ081347.5+494110}
\end{figure*}

\begin{figure*}
\begin{center}
\includegraphics[scale=0.45]{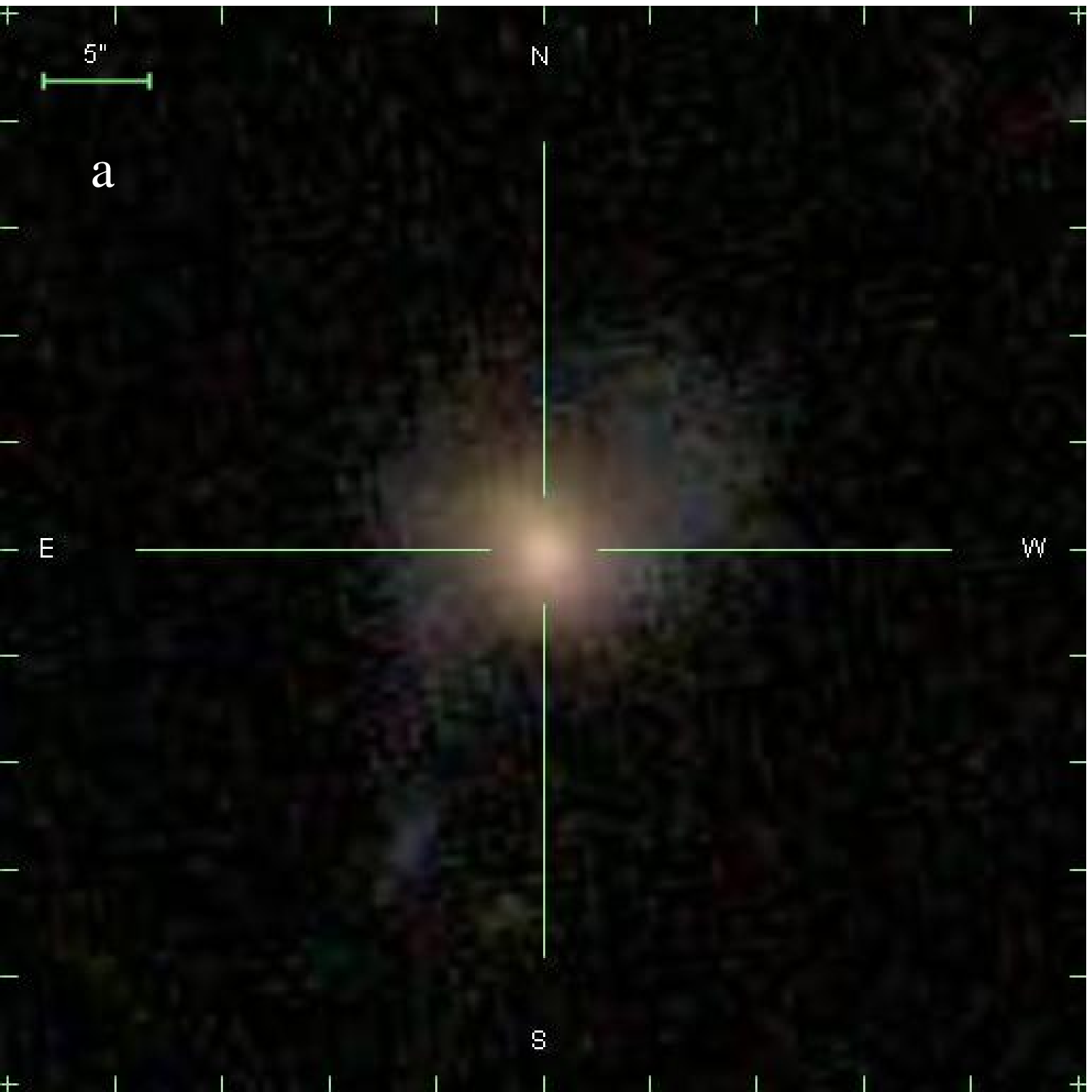}
\includegraphics[scale=0.45]{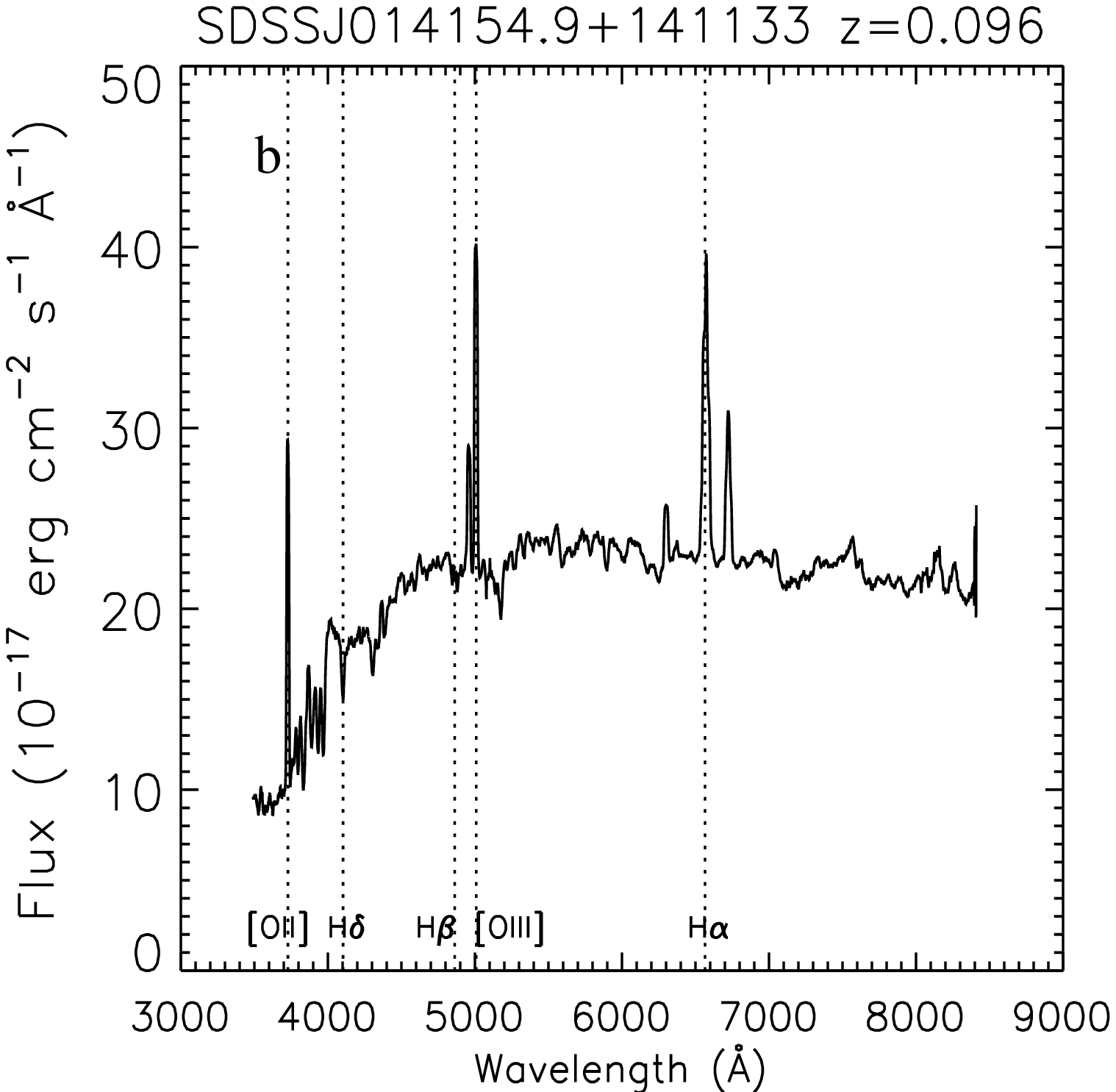}
\includegraphics[scale=0.40]{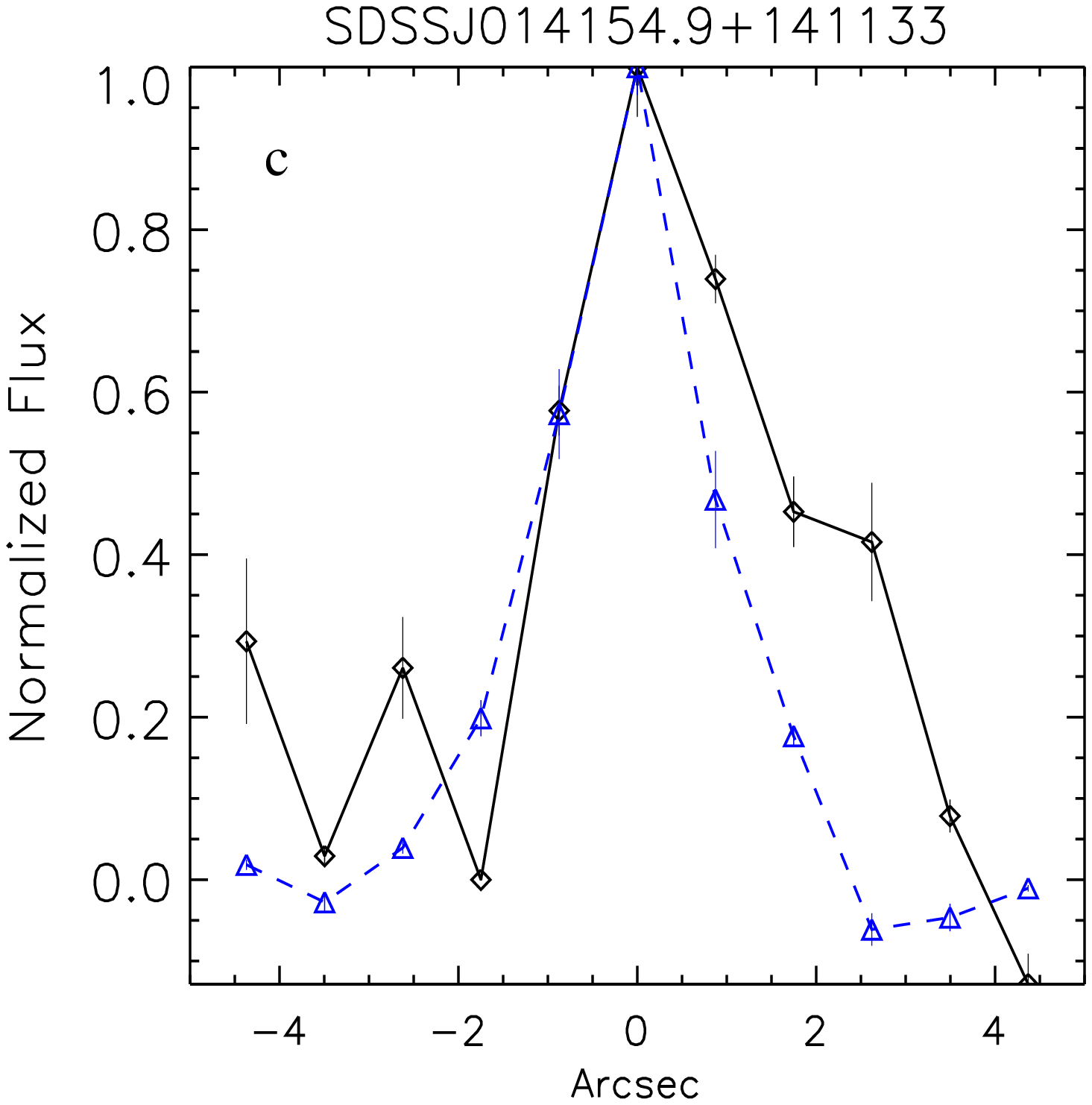}
\includegraphics[scale=0.40]{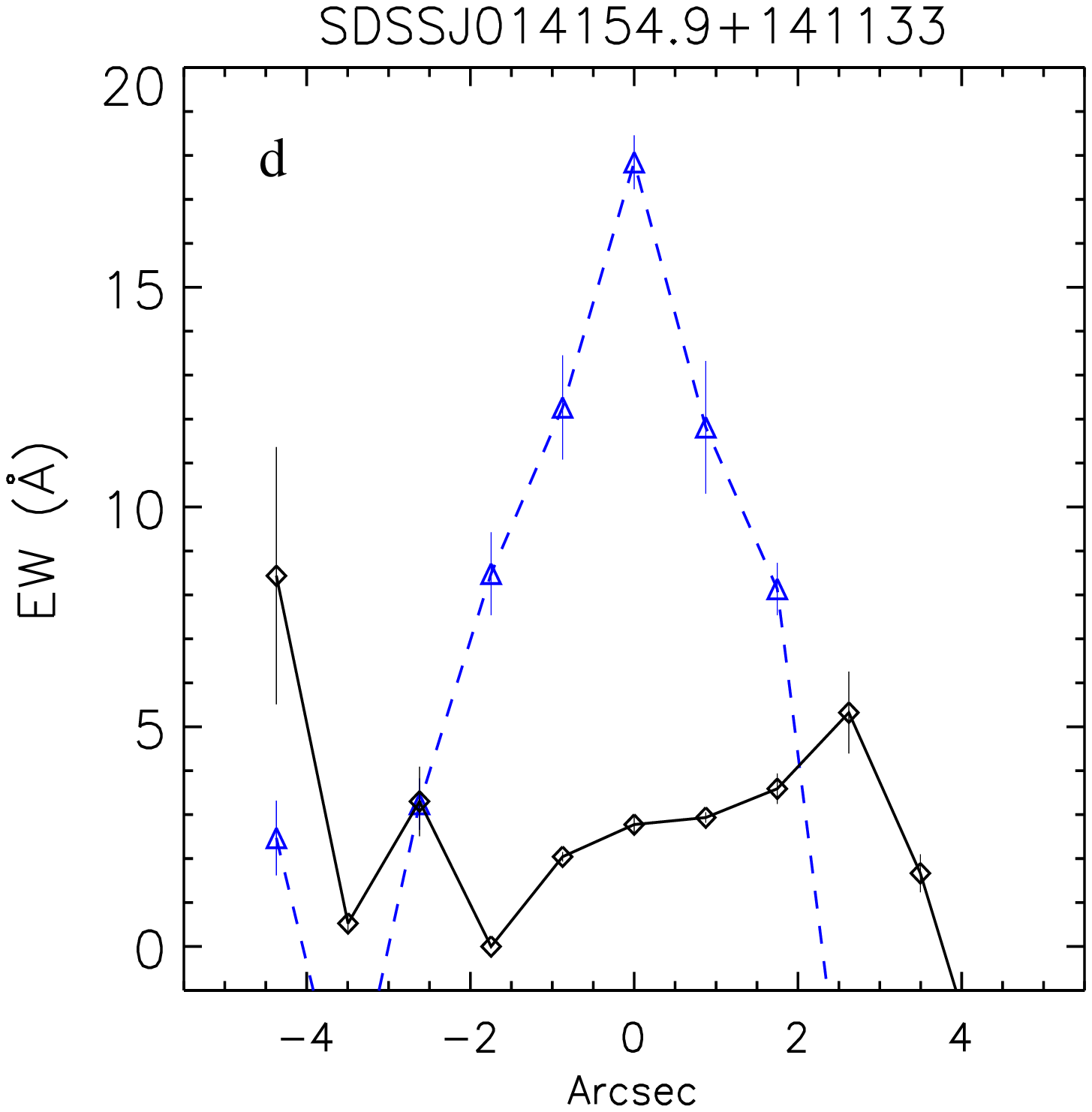}
\includegraphics[scale=0.40]{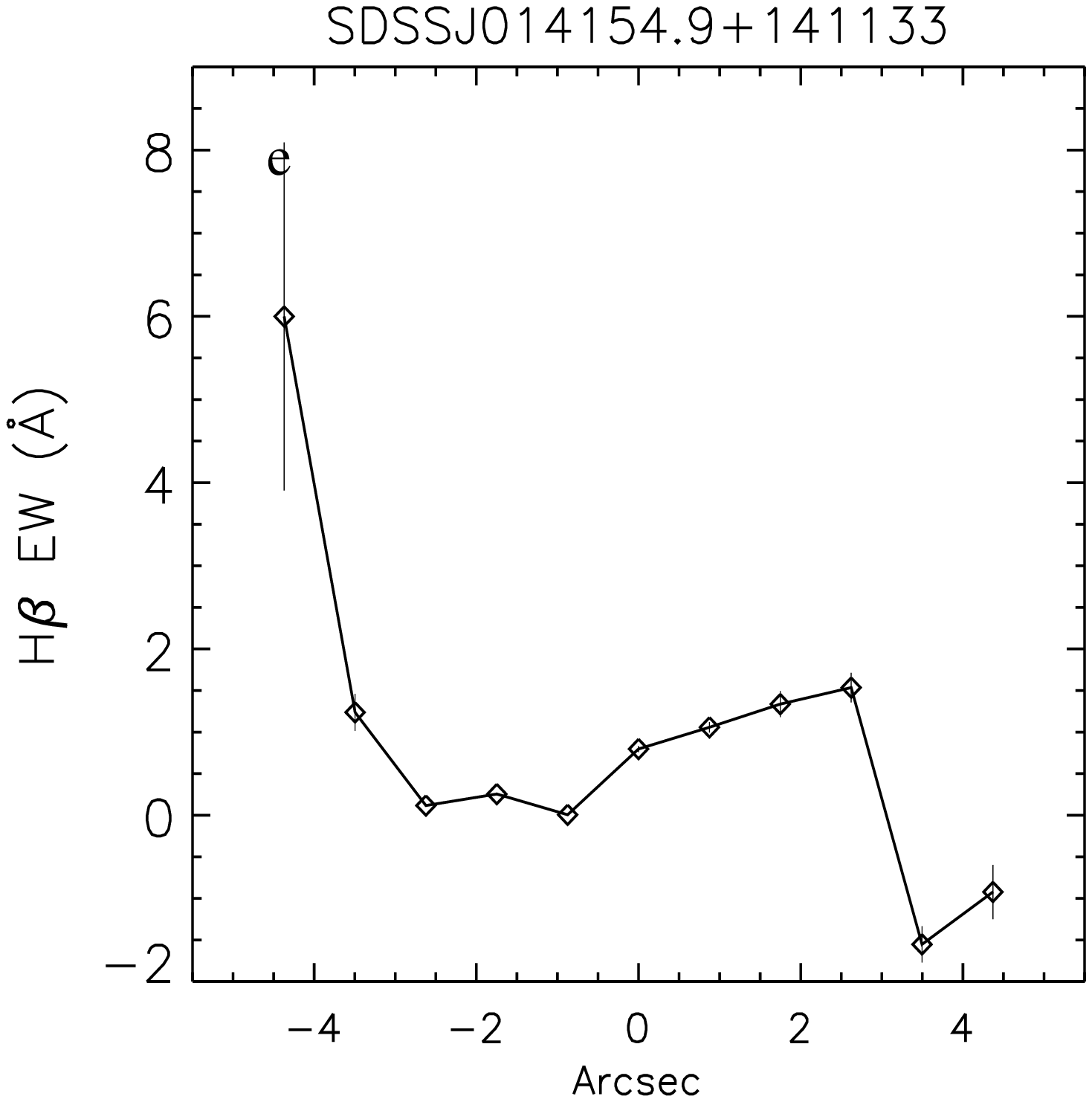}
\includegraphics[scale=0.40]{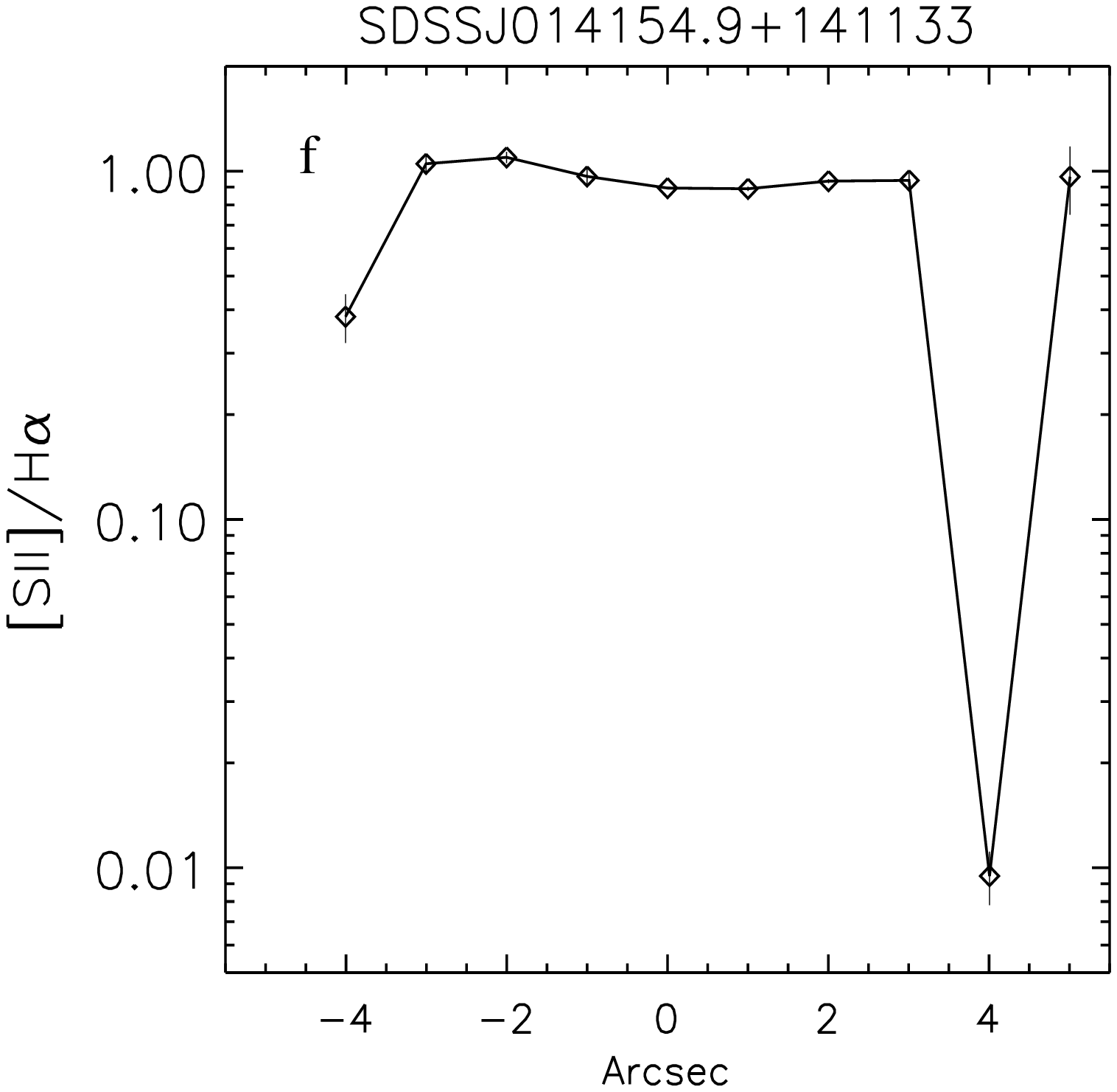}
\end{center}
\caption{As figure \ref{fig:SDSSJ023301.2+002515}, but for SDSSJ014154.9+141133. 
}\label{fig:SDSSJ014154.9+141133}
\end{figure*}


\section{Discussion}\label{discussion}

\subsection{Abundance of HDSAGNs}\label{abundance}

In Section \ref{data}, we identified an interesting population of galaxies, which have both the AGN and post-starburst signatures, suggesting that the AGN may outlive the starburst in the starburst-AGN connection. Here, it is important to investigate how significant this AGN-poststarburst phase is to the overall AGN/starburst evolution in the entire universe. In order to investigate the relative abundance of HDSAGNs, we need to remove a Malmquist bias by constructing a volume-limited sample of galaxies. Since the SDSS main galaxy sample is flux-limited at $r$=17.77 \citep{2002AJ....124.1810S}, we limit our sample to $0.01<z\leq0.09$ and $M_r\leq -20.39$. We used the k-correction of \citet[][v3\_2]{2003AJ....125.2348B} in computing the absolute magnitude. This is a similar criterion that was used in previous work \citep{Goto2003MD,Goto2003PS,2003PhDT.........2G,Goto2005vdisp}, and is optimized to include large number of 47930 galaxies in the volume limited sample. There are 3458 AGNs among this volume limited sample, i.e.,  7.2\% of the whole galaxies in the volume limited sample are AGNs. This fraction is somewhat lower than previous estimates. \citet{2001ApJ...559..606C} reported an AGN fraction of 17\%. \citet{1995ApJS...98..477H} reported that 43\% of galaxies in their sample of 486 nearby bright galaxies could be classified as an AGN. \citet{2003ApJ...597..142M} measured an AGN fractions of $>$20\%. However, our lower fraction is not inconsistent with these previous results, instead, rather reflects our conservative four line criteria in selecting AGNs. \citet{2001ApJ...556..121K} criterion is conservative for AGNs (i.e., the upper limit for pure starburst galaxies) and that we only select AGNs when all the four lines are securely detected. It is important to keep in mind that there could be $\sim$6 times more AGNs if we loosen our selection criteria.
 
Compared with all the galaxies in the volume limited sample, 0.3\% (147 out of 47930) are HDSAGNs. This is a very small fraction, but is similar to the fractions of E+A (post-starburst) galaxies \citep[0.1\%;][]{Goto2003ea,Goto2005266ea}, perhaps reflecting the short life time of the post-starburst phase ($\sim$1 Gyr). 
  Among the 3458 AGNs, we found 147 HDSAGNs, i.e., 4.2\% of AGNs have a post-starburst signature. This fraction is higher than that of E+As to all galaxies (0.2\%), and therefore, may imply that the post-starburst phase occurs more frequently in AGNs than in normal galaxies. Although 4.2\% is a small number in the local universe, if the fraction is constant throughout the history of the universe, a significant fraction of AGNs may have experienced a post-starburst phase.  With the estimated lifetime of HDSAGNs to be $\sim1$ Gyrs, $\sim$60 \% of all the AGNs could have been through the post-starburst phase in the history of the universe. In addition, \citet{2005astro.ph..3401L} reported that the abundance of massive post-starburst galaxies dramatically increased from the present to $z\sim 1.2$. If there was a similar increase in the abundance of post-starburst--AGNs, more AGNs may have been through the post-starburst phase. There has been found a global similarity in the evolution of star formation rate and the QSO luminosity functions \citep[e.g.,][]{1998MNRAS.293L..49B,2003ApJ...598..886U}. And thus, the post-starburst phase may be an important stage in the overall AGN evolution. 

\subsection{Implications to the Starburst-AGN connection}\label{implications}

 
In Section \ref{sec:Results}, we have performed spatially resolved spectroscopy of three HDSAGNs. Although spatial extent of $starbursting$-AGNs has been studied previously \citep[e.g.,][]{2001ApJ...546..845G,2003MNRAS.339..772R,2005MNRAS.356..270C}, it has been difficult to study spatial extent of $poststarburst$-AGNs until the large data set such as the SDSS become available, simply because strong poststarburst-AGNs 
(H$\delta$ EW $>$4 \AA\ ) are rare.
 In Section \ref{sec:Results}, we have found that the post-starburst regions are more extended than the AGNs region, but centrally concentrated around the central AGNs. It is one of our important results that a spatial connection is found between the post-starburst and the AGN, providing further evidence for a causal relationship between starbursts and AGNs. It is also worth noting that we found the post-starburst regions are centrally concentrated, not in a surrounding ring where the starburst-AGN connection may be more indirect. Some previous work suggested that the starburst may be in a ring surrounding the central AGN \citep[e.g.,][]{1990ApJ...365..502R,1998ApJ...500..147S,1999ApJ...512..140H}, which was not the case with the three HDSAGNs studied here. 

 There have been two possible scenarios in explaining the starburst-AGN connection: (i) starburst evolves into central black holes \citep[e.g.,][]{1983ApJ...266..479W}; (ii)  the central AGN activity stimulates circum-nuclear star formation \citep[e.g.,][]{1999A&AS..135..437G}. Although there still involve many unknown parameters/physics such as unknown lifetime of the AGN and the starburst,  the presence of the poststarburst-AGN galaxies may favor the first case since the starburst is ceasing in these galaxies while the AGN is still active. 
 However, details of the model needs to be improved further since in the original scenario \citep{1983ApJ...266..479W} the AGN features originate from accretion onto numerous small black holes and neutron stars that were once part of the starburst. In this scenario, therefore, the spatial distributions of the starburst and the AGN regions are expected to be similar. In this work, however, we observationally found that the post-starburst regions is spatially more extended than the AGN region. 
In order to collect further evidence on the subject, it is useful to directly observe HI gas in 21cm, and/or to trace the evolution of HDSAGNs as a function of an age indicator of the post-starburst phase as proposed by \citet{YamauchiEA}.
 
 It is also important to remember that the above scenarios assume a physical connection between the starburst and the AGN. There still is a third possibility that two phenomena may be indirectly connected, and that both are just triggered by the same gas fueling \citep[e.g.,][]{2001ApJ...558...81C}. In this case, the starburst may have consumed the fueling gas more rapidly than the central AGNs to evolve into the post-starburst phase. If so, as was proposed by \citet{1997ApJ...480L...5C,2001ApJ...555..719C,1999ASPC..192...69G}, the age of the starburst may be used as a clock to measure the age of the AGN activity, which potentially can provide precious constraints on the lifetime and duty cycles of AGNs. \citet{1994ApJ...423L..27Y} found a correlation between X-ray and CO luminosity in nearby Seyferts, concluding that more powerful AGNs live in more actively star-forming host galaxies. \citet{2003MNRAS.346.1055K} found that the hosts of high-luminosity AGN have much younger stellar age. These results may be implying a possibility in using the age of the starburst to clock AGNs.

 It is of considerable interest to investigate the origin of the post-starburst phenomena in HDSAGNs.  
 Possible ways to terminate starburst while the AGN is still active include a simple fuel exhaustion as the gas supply gravitationally accretes inward \citep[e.g.,][]{2001ApJ...558...81C}, and the starburst winds expelling the less dense ISM at larger radii \citep[e.g.,][]{1987A&A...173...23A}. 
 \citet{Goto2005266ea} proposed that the origin of E+A galaxies (post-starburst galaxies without AGNs nor on-going star formation) is likely to be a dynamical interaction/merger with close companion galaxies by finding statistical excess in the number of companion galaxies within 100 kpc. 
 There has been cumulative evidence that the morphology of E+A galaxies is dynamically disturbed \citep[e.g.,][]{2004MNRAS.355..713B,2005MNRAS.359.1421P,YamauchiEA}. 
 There is a possibility that a similar dynamical merger/interaction could be responsible for the starburst-AGN connection or the post-starburst-AGN connection. 
 During such dynamical interaction/merger, interstellar gas experiences strong torques and loss of angular momentum. Consequently, ISM  should be transported toward the centre of the galaxy to feed the nuclear starburst and the central AGN \citep{1988ApJ...332..163S,1988ApJ...332..124N,1991IAUS..144..337N,1996ApJ...471..115B}.  However, in such a case, the merger/interaction has to only terminate the starburst to create a poststarburst-AGN as was found in this study. 
 Observationally,  \citet{2000AJ....119...59C} reported that a companion galaxy of the previously found poststarburst QSO UNJ1025-004 is at the same redshift. 
 In the panels (a) of Figs \ref{fig:SDSSJ023301.2+002515}-\ref{fig:SDSSJ014154.9+141133}, we do not find any sign of strong merger/interactions, but SDSSJ081347.5+494110 is double peaked and SDSSJ014154.9+141133 may have a low surface brightness tidal tail. A higher resolution imaging survey of our HDSAGNs is required to conclude on the subject.

\section{Conclusions}\label{conclusion}

 By taking advantage of the large number of galaxy spectra taken with the SDSS, we have identified 840 galaxies with both the post-starburst signature and the central AGN. The presence of these galaxies themselves suggests that the starburst may outlive AGNs in the starburst-AGN connection. The abundance of these galaxies is small in the local universe (4.2\%), but due to the short lifetime of the post-starburst phase ($\sim$1 Gyr), potentially majority of AGNs ($\sim$60\%) may have been through the phase in the history of the universe. 
 We further performed a spatially resolved long-slit spectroscopy of three HDSAGNs to find that (i) the post-starburst region is  concentrated around the central AGN, but more extended than the AGN region, suggesting the spatial connection between the starburst and AGN; (ii) after deconvolving PSF, the effective radii of the H$\delta$ profile is larger than that of the [OIII], being consistent with the above picture where the post-starburst co-exists with the AGN, but is spatially more extended.  These spatial correlations support a physical connection between the starburst and the AGN, in which the AGN outlives the starburst. 

\section*{Acknowledgments}
 We thank Drs. Masafumi Yagi and Ricardo Demarco for useful discussion, 
and an anonymous referee for many insightful comments that improved the paper.

 This research is based on observations obtained with the Apache Point Observatory 3.5-meter telescope, which is owned and operated by the Astrophysical Research Consortium.
   
    Funding for the creation and distribution of the SDSS Archive has been provided by the Alfred P. Sloan Foundation, the Participating Institutions, the National Aeronautics and Space Administration, the National Science Foundation, the U.S. Department of Energy, the Japanese Monbukagakusho, and the Max Planck Society. The SDSS Web site is http://www.sdss.org/.

    The SDSS is managed by the Astrophysical Research Consortium (ARC) for the Participating Institutions. The Participating Institutions are The University of Chicago, Fermilab, the Institute for Advanced Study, the Japan Participation Group, The Johns Hopkins University, Los Alamos National Laboratory, the Max-Planck-Institute for Astronomy (MPIA), the Max-Planck-Institute for Astrophysics (MPA), New Mexico State University, University of Pittsburgh, Princeton University, the United States Naval Observatory, and the University of Washington.




%
%
%



\end{document}